\documentclass[12pt,a4paper]{article}
\usepackage[T2A]{fontenc}
\usepackage[utf8]{inputenc}
\usepackage[english]{babel}
\usepackage{amssymb}
\usepackage{amsmath}
\usepackage{graphicx}
\usepackage[small,sc,center]{titlesec}
\usepackage{indentfirst}
\usepackage{siunitx}
\usepackage[labelsep=period]{caption}
\usepackage{caption}
\begin{document}

\begin{center}
	\textbf{Nonlinear models of Mira variables with long secondary periods}

	\vskip 3mm
	\textbf{Yu. A. Fadeyev${}^1$\footnote{E--mail: fadeyev@inasan.ru}}

\vskip 12pt
\textit{$^1$Institute of Astronomy, Russian Academy of Sciences, Moscow, 119017 Russia} \\

	Received October 10, 2026; revised October 22, 2026; accepted October 25, 2026
\end{center}

\vskip 10pt
\noindent
\textbf{Abstract} ---
Calculations of nonlinear radial oscillations in asymptotic giant branch stars with an initial
mass of $M_\mathrm{ZAMS}=1.6M_\odot$ were performed for initial helium and metal abundances of
$Y=0.28$ and $Z=0.006$, respectively.
Several models from the evolutionary sequence after the 4th thermal pulse were used as initial
conditions for the hydrodynamic equations describing stellar pulsations with long secondary periods.
Using the discrete Fourier transform of the hydrodynamic solutions we show that long secondary
periods appear in Mira variables pulsating simultaneously in the fundamental mode and first
overtone with non-multiple periods.
The relative deviation from multiplicity is $\delta_\mathrm{\Pi}\le 1.2\times 10^{-2}$ and is
responsible for mode splitting into equally spaced frequency components, each of which is
a standing wave with a free outer boundary.
The cycle length of long secondary periods is nearly 20 times that of the fundamental mode.
Long secondary periodicity is a nonlinear effect that arises after reaching the limiting amplitude
and develops over several hundred fundamental mode periods.


\newpage
\section{INTRODUCTION}

Long secondary periods in Mira--type pulsating variables were discovered (Wood et al., 1999)
through the light curve analysis of nearly a thousand red giants in the Large Magellanic Cloud (LMC)
observed over nearly six years within the framework of the MACHO project.
In particular, about a quarter of the observed stars were found to exhibit long secondary periods
(LSPs) with cycle lengths an order of magnitude longer than the fundamental mode period.
Later, these conclusions were confirmed by more extensive photometric data from
the EROS (Cioni et al. 2001) and OGLE (Soszy{\'n}ski et al., 2004) surveys.

Although the LSP was discovered decades ago, its underlying nature remains unclear.
The difficulty arises primarily because the maximum period of acoustic waves in a radially
pulsating star is constrained by the fundamental mode, which is much shorter than the long
secondary period timescale.
The models proposed to explain the origin of the LSP consider nonradial stellar oscillations
(Wood et al. 2004), the presence of a low--mass orbiting companion
(Wood et al. 2004; Soszy{\'n}ski 2007; Soszy{\'n}ski et al. 2021),
and the turnover of giant convection cells (Stothers 2010; Saio et al. 2015).
Unfortunately, all these models are thought to fail, and the problem of the LSP remains
unsolved.
A more detailed discussion of the models proposed to explain the origin of the LSP is given
in the review paper by Skowron and Soszy{\'n}ski (2026).

In his previous study of nonlinear pulsations of Mira variables in the LMC the author (Fadeyev 2024)
showed that some models exhibit the LSP phenomenon with cycle lengths 10 -- 30 times longer than
the fundamental mode period.
Additional computations revealed that the LSP cycle length does not depend on the spatial or
temporal grid of the hydrodynamic model and is determined by the mass and age of the star.

The goal of the present study is to determine the origin of the LSP in pulsating red giants.
To this end, we computed a dozen Mira hydrodynamic models with close values of the star age
near the point on the evolutionary track where regular stellar oscillations transform into the LSP.
The results presented below were obtained from consistent stellar evolution and nonlinear stellar
pulsation calculations, where the equations of radiation hydrodynamics and time--dependent convection
are solved with initial conditions determined from models of the evolutionary sequence.
The computational methods are discussed in Fadeyev (2013).

\section{RESULTS OF COMPUTATIONS}

The initial conditions required for the solution of the hydrodynamic equations describing nonlinear stellar
oscillations were calculated on the basis of the evolutionary sequence with an initial mass of
$M_\mathrm{ZAMS}=1.6M_\odot$ and an initial helium abundance of $Y=0.28$.
Similar to the previous study (Fadeyev 2024), the evolutionary calculations were performed with an initial
metal abundance of $Z=0.006$, corresponding to the mean metal abundance in the LMC (Cioni et al. 2006).

\begin{figure}
 \centering
 \includegraphics[width=0.9\columnwidth,clip]{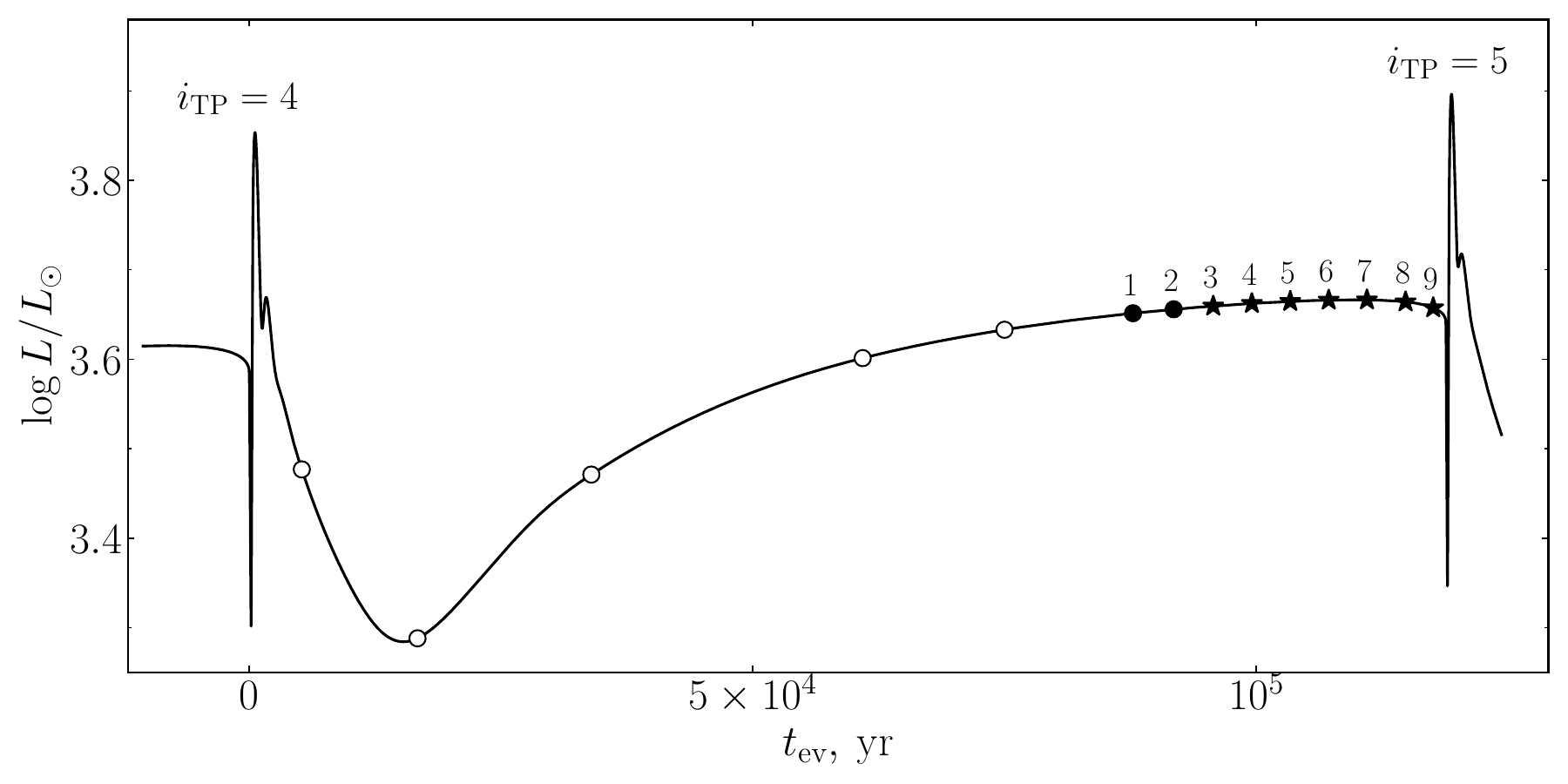}
 \caption{Variations of stellar luminosity between the fourth ($i_\mathrm{TP}=4$) and fifth
          ($i_\mathrm{TP}=5$) thermal pulses of the helium-burning shell.
          Hydrodynamic models with decaying oscillations are shown by open circles.
          Models with regular oscillations and those with the LSP are marked by filled circles
          and asterisks, respectively.}
\label{fig1}
\end{figure}

In the models of the computed evolutionary sequence, the LSP appears after the fourth thermal pulse
of the helium-burning shell.
The evolutionary status of the hydrodynamic models is illustrated in Fig.~\ref{fig1} by a plot
of stellar luminosity variations over the time interval $\Delta t_\mathrm{ev}=1.17\times 10^5$~yr
between the fourth and fifth thermal pulses.
For the sake of convenience, the evolutionary time $t_\mathrm{ev}$ is set to zero at the maximum
energy release by the helium-burning shell at $i_\mathrm{TP}=4$, when the stellar mass and age
are $1.48\,M_\odot$ and $1.595$~Gyr, respectively.

Open circles in Fig.~\ref{fig1} mark the hydrodynamic models with decaying kinetic energy of
oscillations.
Filled circles represent the models with regular limit-cycle pulsations, whereas asterisks indicate
the models exhibiting the LSP.
The model numbers shown in Fig.~\ref{fig1} will be used in the further discussion.

\begin{figure}
 \centering
 \includegraphics[width=0.9\columnwidth,clip]{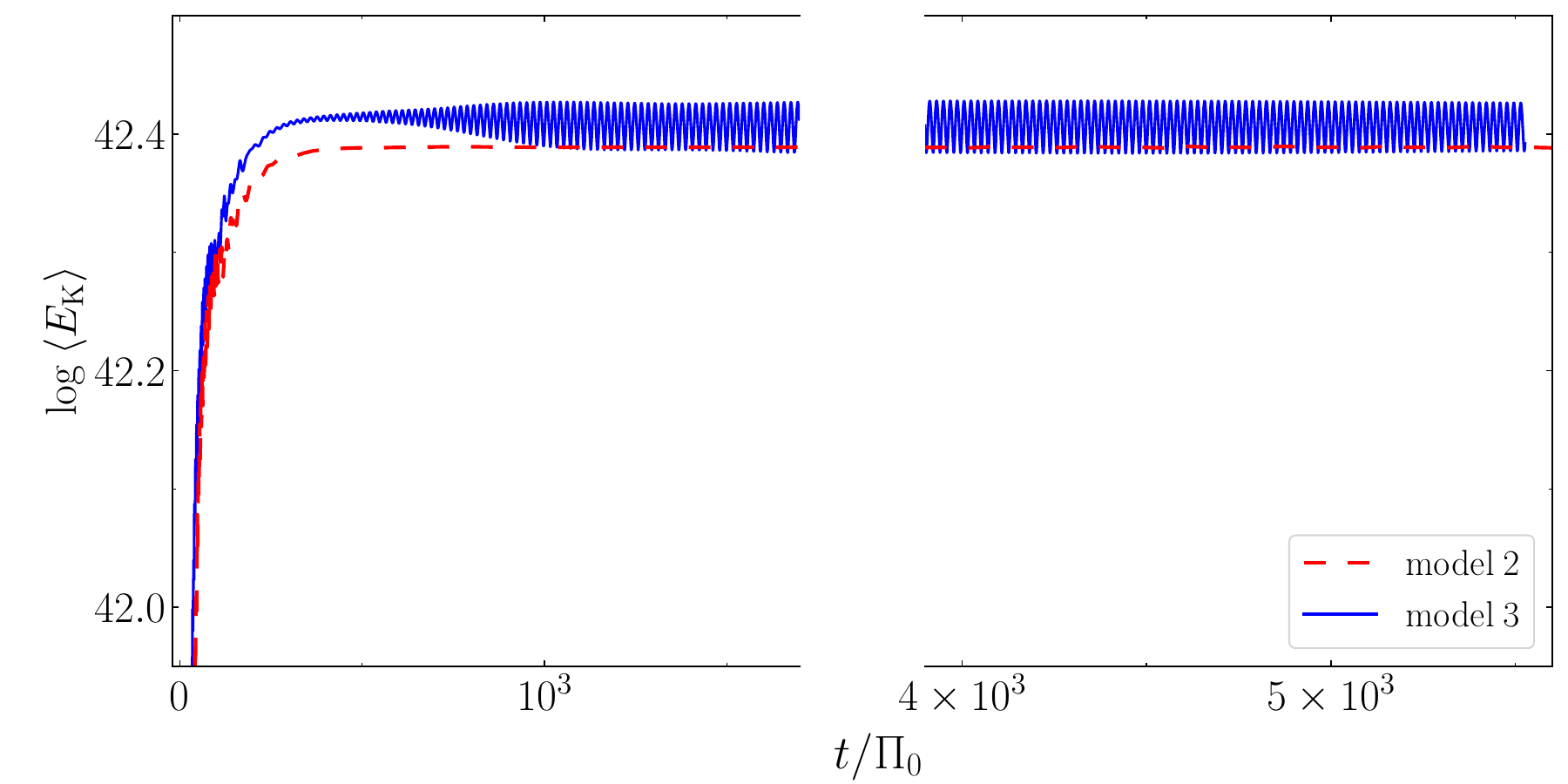}
 \caption{Variations of the mean kinetic energy (averaged over the pulsation period)
          in hydrodynamic models 2 and 3.}
 \label{fig2}
\end{figure}

The solution of the hydrodynamic equations describes self-excited stellar oscillations that
grow from small perturbations appearing due to the transformation of the mesh variables from
the evolutionary model to the Lagrangian grid of the hydrodynamic model.
Fig.~\ref{fig2} shows temporal variations of the kinetic energy averaged over a pulsation period
$\langle E_\mathrm{K}\rangle$ in hydrodynamic models 2 and 3 separated in stellar age by
$\approx 4\times 10^3$~yr.
It should be noted that the time $t$ in these plots is the independent variable of the Cauchy problem
for the equations of hydrodynamics and, therefore, is irrelevant to the stellar evolution problem.
As seen in Fig.~\ref{fig2}, the cessation of amplitude growth in model 2 is followed by the
attainment of limit-cycle oscillations, which is due to saturation of the $\kappa$-mechanism
driving stellar oscillations.
In contrast, the plot for model 3 shows an amplitude increase in long secondary variations of
$\langle E_\mathrm{K}\rangle$ during limit-cycle oscillations.
The appearance of long secondary variations, as well as the saturation of the $\kappa$-mechanism,
is a nonlinear effect, and its growth timescale is as long as several hundred fundamental-mode periods
$\Pi_0$.

\begin{figure}
 \centering
 \includegraphics[width=0.9\columnwidth,clip]{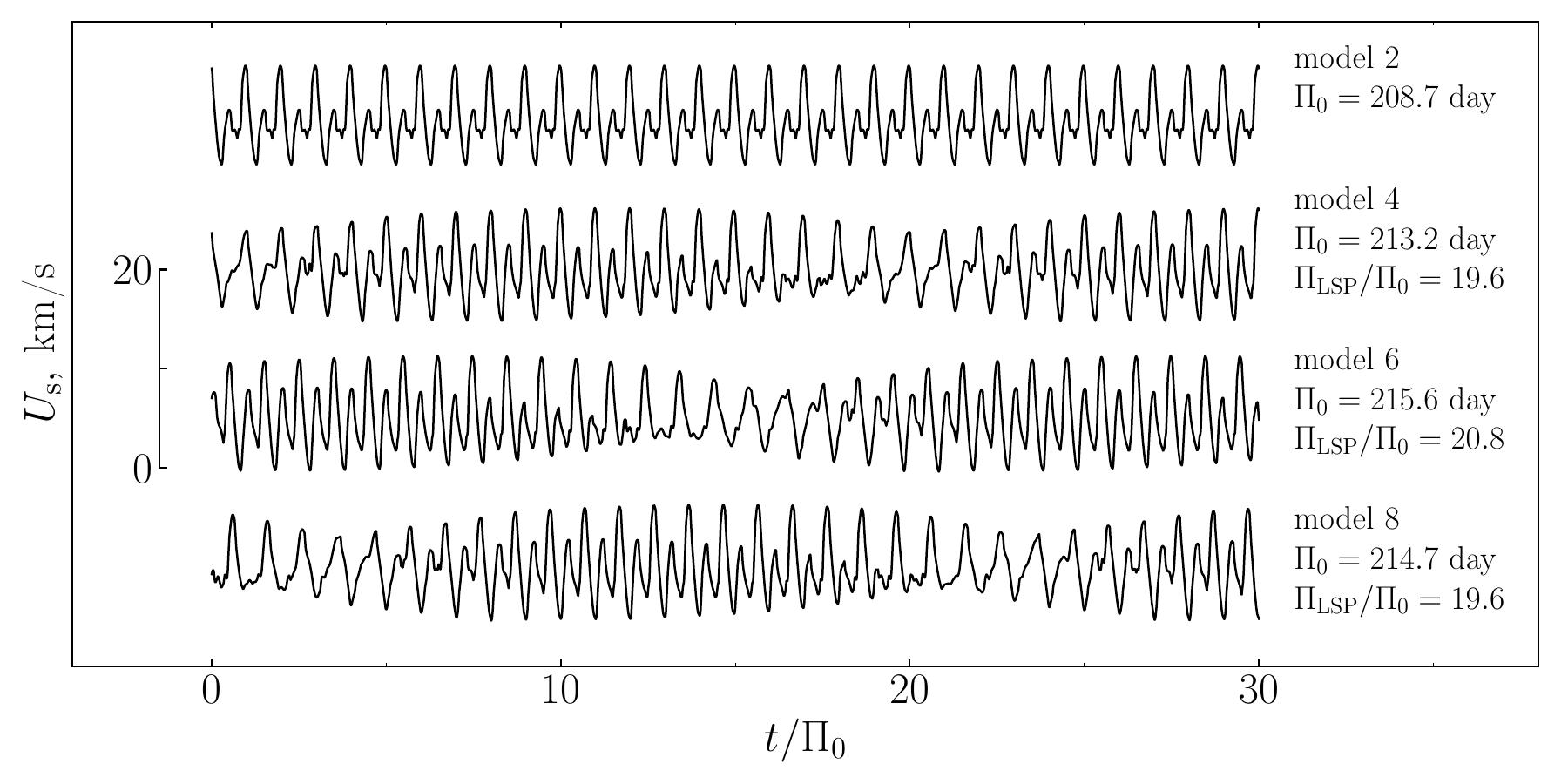}
 \caption{Temporal variations of the gas flow velocity at the outer boundary of hydrodynamic models.}
\label{fig3}
\end{figure}

The second periodicity during limit-cycle oscillations is depicted in Fig.~\ref{fig3}, which shows
the temporal variations of the gas flow velocity $U_\mathrm{s}$ at the outer boundary for several
hydrodynamic models.
For convenience, the zero point of the independent variable $t$ in Fig.~\ref{fig3} is chosen
arbitrarily because the plots correspond to the stage of limit-cycle oscillations.
For models 4, 6, and 8 the ratio of the LSP cycle length to the fundamental-mode period is
$\Pi_\mathrm{LSP}/\Pi_0\approx 20$.

\begin{figure}
 \centering
 \includegraphics[width=0.9\columnwidth,clip]{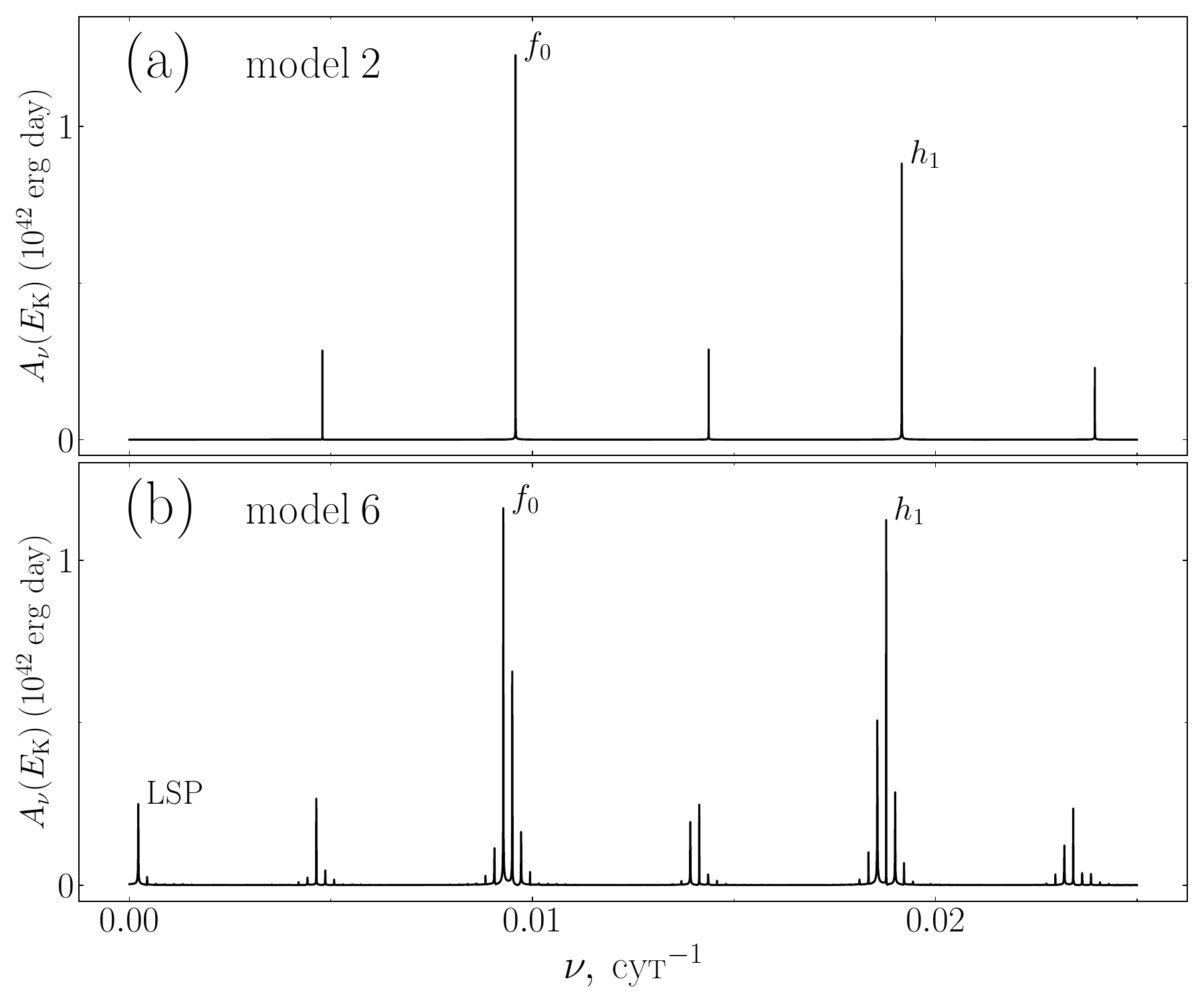}
 \caption{Spectral density of the kinetic energy in hydrodynamic models 2 (a) and 6 (b).
          The labels $f_0$, $h_1$, and LSP correspond to the frequencies of the fundamental mode,
          the first overtone, and the long secondary period.}
\label{fig4}
\end{figure}

Alternating amplitudes in velocity variations of model 2 (see Fig.~\ref{fig3}) allow us to conclude
that stellar pulsations are represented by both the fundamental mode and the first overtone.
Indeed, the spectral density of the kinetic energy in model 2 (see Fig.~\ref{fig4}a) is characterized
by two almost equal peaks corresponding to the fundamental mode and the first overtone with
commensurate periods: $\Pi_0 = 208.75067$~day and $\Pi_1 = 104.37534$~day, that is, $\Pi_0/\Pi_1 = 2$.
It should be noted that in radial stellar pulsations in the form of standing waves,
the frequency of the kinetic energy $E_\mathrm{K}$ is twice the frequency of stellar pulsations:
$\nu_\mathrm{EK} = 2/\Pi_0$.

However, the situation is different in hydrodynamic models exhibiting secondary periods.
As seen in Fig.~\ref{fig4}b, the peak $f_0$ in the spectral density of the kinetic energy of model 6
corresponds to the period of the fundamental mode $\Pi_0=215.58779$~day, which is not a multiple
of the first overtone period $\Pi_1 = 106.51607$~day, that is, $\Pi_0/\Pi_1 \approx 2.024$.
Therefore, the secondary periodicity appears due to amplitude variations produced by the superposition
of two harmonics with slightly different frequencies.

Of great interest are also the peaks that appear near the main frequency maxima $f_0$ and $h_1$ in the
LSP hydrodynamic models (see Fig.~\ref{fig4}b).
In all hydrodynamic models with secondary periods, the frequency interval between the main maximum
and each peak is given by $\delta\nu = \pm i\Delta\nu$, where $i$ is the order of the peak and
$\Delta\nu$ is a constant determined by the stellar structure.

\begin{figure}
 \centering
 \includegraphics[width=0.7\columnwidth,clip]{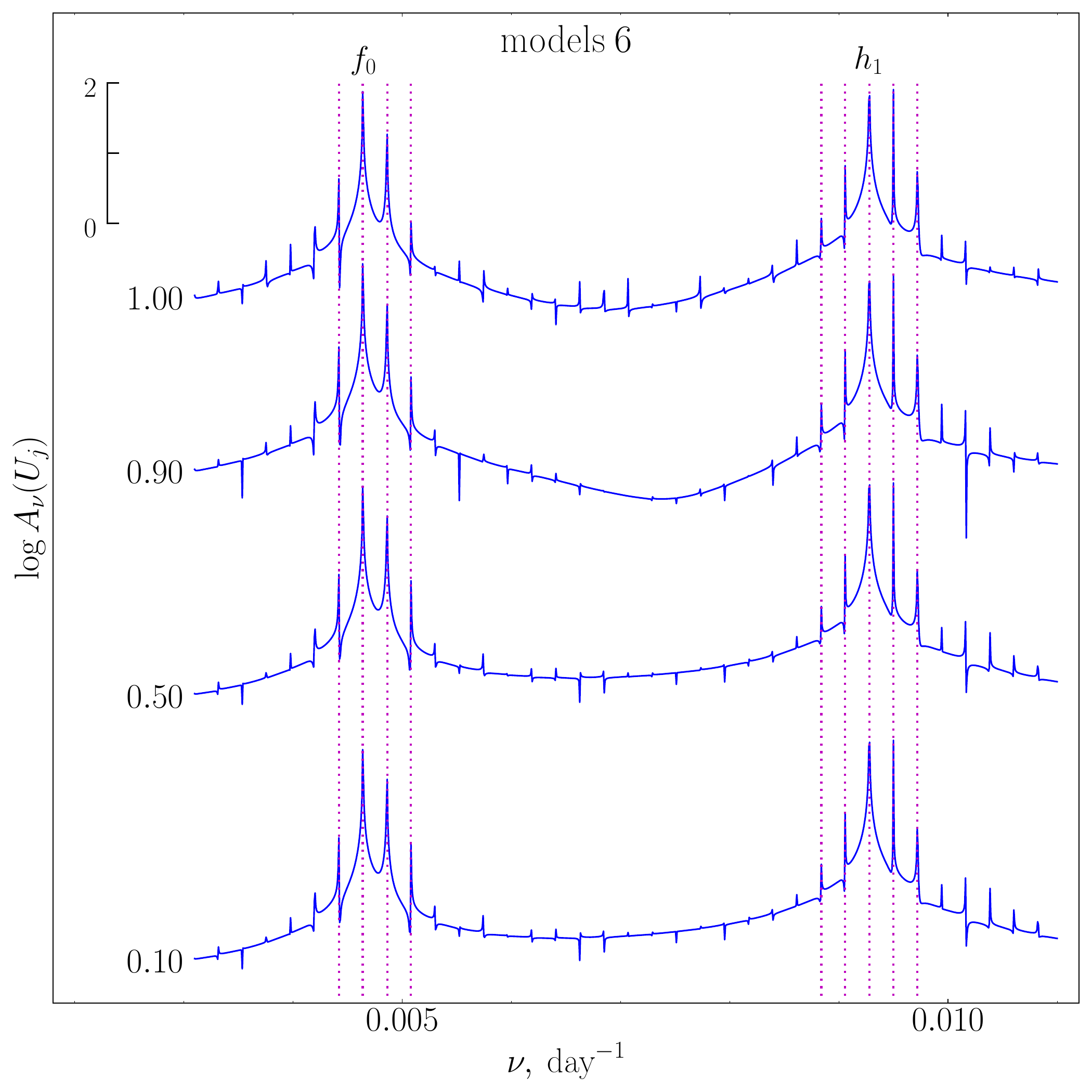}
 \caption{Spectral density of the gas flow velocity in several mass zones of hydrodynamic
          model 6 (solid lines).
          Values of the average radius $\langle r_j\rangle / \langle r_N\rangle$ of the zones are given
          near the plots.
          Vertical dotted lines mark the frequency maxima of the split modes.}
\label{fig5}
\end{figure}

\begin{figure}
 \centering
 \includegraphics[width=0.9\columnwidth,clip]{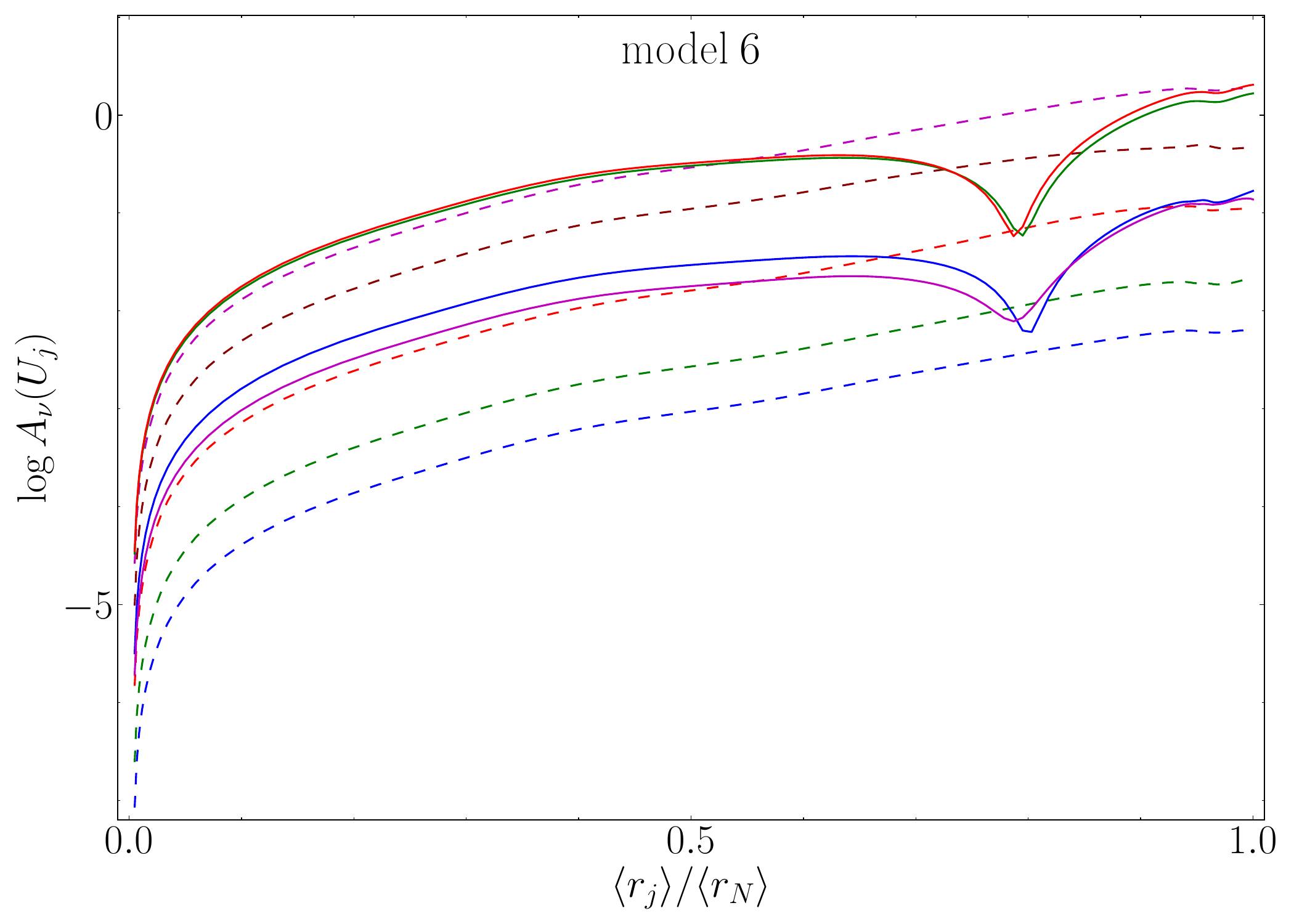}
 \caption{Radial distribution of the spectral density of the gas flow velocity at the frequencies
          of the split fundamental mode (dashed lines) and the split first overtone (solid lines).}
\label{fig6}
\end{figure}

The origin of the secondary periodicity can be better understood from the plots of the spectral
density of the gas flow velocity $U_j$ in several mass zones of hydrodynamic model 6 shown in
Fig.~\ref{fig5}.
The mean values of the radii of these zones lie within the range
$0.1\le\langle r_j\rangle/\langle r_N\rangle\le 1$
and are given next to each plot.
As can be seen, the frequencies of the split fundamental mode $f_0$ and the split first overtone $h_1$
do not change throughout the entire pulsating stellar envelope and, therefore, represent
standing waves.
The plots of the radial distribution of the spectral density of the gas flow velocity $U_j$ at
the frequencies marked by dotted lines in Fig.~\ref{fig5} (see Fig.~\ref{fig6}) allow us to conclude
that the side peaks represent the fundamental mode and the first overtone.

The main characteristics of the hydrodynamic models of Mira variables are listed in Table~\ref{table1},
where the model number is given in the first column.
The evolutionary time $t_\mathrm{ev}$ in the second column is counted from the maximum of $L_\mathrm{He}$
at the fourth thermal pulse ($i_\mathrm{TP}=4$), when the stellar age was 1.594453~Gyr.
In the third and fourth columns, we list the stellar mass $M$ and the stellar bolometric luminosity $L$.
It should be noted that the luminosity $L$ is one of the inner boundary conditions and was also used
to control the accuracy of hydrodynamic computations.
In particular, we apply the criterion $\left| 1 - \langle L_\mathrm{s}\rangle / L\right| < 10^{-3}$,
where $\langle L_\mathrm{s}\rangle$ is the luminosity at the outer boundary averaged over
the limit-cycle oscillation interval.
In the fifth column, we give the stellar radius $\langle R\rangle$ averaged over the pulsation period,
whereas in the sixth and seventh columns, we list the periods of the fundamental mode $\Pi_0$ and
the first overtone $\Pi_1$, determined from the frequency maxima of the spectral density of
the stellar pulsation kinetic energy.
In the eighth column, we list the values of the relative deviation from the harmonic relationship
$$
\delta\Pi_{01} = \left|1 - \dfrac{1}{2}\dfrac{\Pi_0}{\Pi_1}\right| ,
$$
whereas in the ninth column, we give the ratio of the LSP cycle length to the
fundamental-mode period $\Pi_\mathrm{LSP}/\Pi_0$.

\section{CONCLUSIONS}

The results presented in this paper allow us to conclude that to explain the origin of the LSP
in Mira variables there is no need to invoke hypotheses implying the absence of spherical symmetry.
As is known from the theory of oscillations, the superposition of two harmonics with a small
frequency difference leads to frequency splitting and beats with a longer period (Smith 2020).
In Mira variables, the role of harmonic oscillators belongs to the fundamental mode and the
first overtone.
For a relative deviation from the harmonic relationship $\delta\Pi_{01}\lesssim 2\%$, both
harmonics split, and their frequency satellites are standing waves with a free outer boundary,
whereas the resulting beats play the role of the LSP.
This conclusion is based on the results of the consistent stellar evolution and pulsation
computations, and thus the proposed model of the LSP supports the theory of the late stages of
stellar evolution.

For a more detailed comparison of the theory with observations it is necessary to undertake significantly
more extensive computations of both stellar evolution and nonlinear pulsations.
The main difficulty of this problem is that there is not yet a general recipe for
predicting the stellar age and mass corresponding to the physical conditions in the star
when the fundamental mode and the first overtone experience frequency splitting.

\section*{REFERENCES}

\begin{enumerate}

 \item M.-R.L. Cioni, J.-B.~Marquette, C.~Loup, M.~Azzopardi, H.J.~Habing, T.~Lasserre, and E.~Lesquoy,
       ``Variability and spectral classification of LMC giants: Results from DENIS and EROS'',
       Astron. Astrophys. \text{377}, 945 (2001).

 \item M.-R.L. Cioni, L.~Girardi, P.~Marigo, and H.J.~Habing,
       ``AGB stars in the Magellanic Clouds. II. The rate of star formation across the LMC'',
       Astron. Astrophys. \text{448}, 77 (2006).

 \item Yu.A.~Fadeyev,
       ``Evolution and the period--luminosity relation for red supergiants in the Magellanic Clouds'',
       Astron. Lett. \textbf{39}, 306 (2013).

 \item Yu.A.~Fadeyev,
       ``Models for Mira variables of the Large Magellanic Cloud'',
       Astron. Lett. \textbf{50}, 561 (2024).

 \item H.~Saio, P.R.~Wood, M.~Takayama, and Y.~Ita,
       ``Oscillatory convective modes in red giants: a possible explanation of the long secondary periods'',
       MNRAS \textbf{452}, 3863 2015).

 \item D.M.~Skowron and I.~Soszy{\'n}ski,
       ``The role of long-period variable stars in observational astrophysics'',
       Ann. Rev. Astron. Astrophys. \textbf{64}, 103 (2026).

 \item W.F.~Smith, \textit{Waves and oscillations} (Oxford University Press, 2010).

 \item I.~Soszy{\'n}ski,
       ``Long secondary periods and binarity in red giant stars'',
       Astrophys. J. \textbf{660}, 1486 (2007).

 \item I.~Soszy{\'n}ski, A.~Udalski, M.~Kubiak, M.K.~Szyma\'nski, G.~Pietrzy\'nski,
       {\.Z}ebru\'n K., O.~Szewczyk, {\L}.~Wyrzykowski., and W.A.~Dziembowski,
       ``The optical gravitational lensing experiment. Ellipsoidal variability of red giants in
       the Large Magellanic Cloud'',
       Acta Astron. \textbf{54}, 347 (2004).

 \item I.~Soszy{\'n}ski, A.~Olechowska, M.~Ratajczak, P.~Iwanek, D.M.~Skowron, P.~Mr\'oz P.,
       P.~Pietrukowicz, A.~Udalski, M.K.~Szyma\'nski, J.~Skowron, M.~Gromadzki, R.~Poleski, S.~Koz{\l}owski,
       M.~Wrona, K.~Ulaczyk, and K.~Rybicki,
       ``Binarity as the origin of long secondary periods in red giant stars'',
       Astrophys. J. \textbf{911}, L22 (2021).

 \item R.B.~Stothers,
       ``Giant convection cell turnover as an explanation of the long secondary periods in semiregular
       red variable stars'',
       Astrophys. J. \textbf{725}, 1170 (2010).

 \item P.R.~Wood, C.~Alcock, R.A.~Allsman, D.~Alves, T.S.~Axelrod, A.C.~Becker, D.P.~Bennett,
       K.H.~Cook, A.J.~Drake, K.C.~Freeman, K.~Griest, L.J.~King, M.J.~Lehner, S.L.~Marshall, G.~Minniti,
       B.A.~Peterson, M.R.~Pratt, P.J.~Quinn, C.W.~Stubbs, W.~Sutherland, A.~Tomaney, T.~Vandehei T., and
       D.L.~Welch,
       ``MACHO observations of LMC red giants: Mira and semi--regular pulsators, and contact and
       semi--detached binaries'',
       IAU Symp. \textbf{191}, 151 (1999).

 \item P.R.~Wood, E.A.~Olivier, and S.D.~Kawaler,
       ``Long secondary periods in pulsating asymptotic giant branch stars: an investigation of their origin'',
       Astrophys. J. \textbf{604}, 800 (2004).

\end{enumerate}

\begin{table}
\caption{Nonlinear models of Mira variables}
\label{table1}
\begin{center}
\begin{tabular}{ccccccccc}
\hline
    & $t_\mathrm{ev}, \textrm{Myr}$ & $M/M_\odot$ & $\log L/L_\odot$ & $\log\langle R\rangle/R_\odot$ &
      $\Pi_0, \textrm{day}$ & $\Pi_1, \textrm{day}$ & $\log\:\delta\Pi_{01}$ & $\Pi_\mathrm{LSP}/\Pi_0$ \\
\hline
 1  &   0.0878  &   1.476  &   3.651  &   2.339  &   204.852  &   102.426  &          &       \\
 2  &   0.0918  &   1.475  &   3.656  &   2.344  &   208.746  &   104.373  &          &       \\
 3  &   0.0957  &   1.474  &   3.660  &   2.347  &   211.780  &   105.898  &   -4.11  &  18.6 \\
 4  &   0.0996  &   1.474  &   3.662  &   2.349  &   213.227  &   106.594  &   -3.75  &  19.6 \\
 5  &   0.1034  &   1.473  &   3.665  &   2.350  &   214.203  &   105.750  &   -1.89  &  19.5 \\
 6  &   0.1072  &   1.472  &   3.666  &   2.352  &   215.588  &   106.516  &   -1.92  &  20.8 \\
 7  &   0.1110  &   1.471  &   3.666  &   2.353  &   215.675  &   106.617  &   -1.94  &  21.7 \\
 8  &   0.1148  &   1.470  &   3.664  &   2.351  &   214.731  &   106.033  &   -1.90  &  19.6 \\
 9  &   0.1176  &   1.470  &   3.658  &   2.346  &   211.041  &   105.500  &   -3.71  &  18.7 \\
\hline
\end{tabular}
\end{center}
\end{table}

\newpage

\end{document}